\documentclass{aastex631}

\usepackage{bm}
\usepackage{natbib}
\usepackage{mathtools}
\usepackage{amsmath}
\usepackage{multirow}
\usepackage{graphicx}
\usepackage{hyperref}

\submitjournal{ApJ}

\shorttitle{Slow Oscillations in Stellar Loops}
\shortauthors{Lim et al.}

\graphicspath{{./}{figures/}}

\begin{document}

\title{Slow Magnetoacoustic Oscillations in Stellar Coronal Loops}

\correspondingauthor{Valery M. Nakariakov}
\email{V.Nakariakov@warwick.ac.uk}

\author[0000-0001-9914-9080]{Daye Lim}
\affiliation{Department of Astronomy and Space Science, Kyung Hee University, Yongin 17104, Republic of Korea}

\author[0000-0001-6423-8286]{Valery M. Nakariakov}
\affiliation{Centre for Fusion, Space and Astrophysics, Physics Department, University of Warwick, Coventry CV4 7AL, UK}
\affiliation{School of Space Research, Kyung Hee University, 1732, Yongin 17104, Republic of Korea}

\author[0000-0001-6216-6944]{Yong-Jae Moon}
\affiliation{School of Space Research, Kyung Hee University, 1732, Yongin 17104, Republic of Korea}
\affiliation{Department of Astronomy and Space Science, Kyung Hee University, Yongin 17104, Republic of Korea}


\begin{abstract}
Slow magnetoacoustic oscillations in stellar coronal loops with gravitational stratification are analyzed with a numerical solution of the boundary-value problem for eigenvalues and eigen functions. In this study, we only focus on the resonant periods. The effects of the gravitational stratification, star mass, loop temperature, and loop length on the properties of slow magnetoacoustic oscillations are investigated. It is shown that the discrepancy between stratified and non-stratified loops is higher in density perturbations than velocity perturbations. When the star has larger mass, higher coronal temperature, and longer loop, the density perturbations in stratified loop are significantly different from the harmonic functions. The periods in the stratified loop are slightly longer than in the non-stratified loop. The periods calculated in our model (14--644 min) are consistent with the periods of stellar quasi-periodic pulsations observed in both soft x-rays (2--70 min) and white lights (8--390 min).  
\end{abstract}


\section{Introduction} \label{sec:intro}
Light curves of solar and stellar flares often have flux modulations called quasi-periodic pulsations (QPPs, see, e.g., \citealt{2020STP.....6a...3K}). In solar flares, QPP periods range from sub-seconds to a few tens of minutes, while in more long-durational stellar flares, QPPs periods reach several hours \citep{2019ApJ...884..160V, 2021SSRv..217...66Z}. QPPs could be produced by different mechanisms, which suggests the existence of several different QPP classes \citep[see, e.g.][]{2019PPCF...61a4024N, 2021SSRv..217...66Z}. Mechanisms responsible for QPPs in solar flares could be broadly classified into three categories: magnetohydrodynamic (MHD) oscillations, repetitive regimes of magnetic reconnection (magnetic dripping), and periodically induced magnetic reconnection by external MHD oscillations \citep{2018SSRv..214...45M}. However, other possibilities remain open too. 

Long-period rapidly-decaying oscillations of a Doppler shift of a coronal emission line have been observed in hot coronal loops with \citet{2002ApJ...574L.101W}. Oscillations of this kind are often called SUMER oscillations, after the instrument used for their detection, the Solar Ultraviolet Measurement of Emitted Radiation (SUMER) on the Solar and Heliospheric Observatory. Traditionally, the term, SUMER oscillations, describes long-period  with the period longer than the expected acoustic transit time along a loop, rapidly-decaying with the damping time about the oscillation period, oscillations of the thermal emission. SUMER class QPP is one of several distinct types of QPP, which are still to be rigorously classified \citep[see the discussion in][]{2019PPCF...61a4024N}. Oscillations of the SUMER type have been also observed with other instruments such as the Bragg Crystal Spectrometer on Yohkoh \citep{2005ApJ...620L..67M, 2006ApJ...639..484M}, the Hinode/EUV Imaging Spectrometer \citep{2008ApJ...681L..41M, 2010NewA...15....8S}, in the imaging data with the Atmospheric Imaging Assembly (AIA) on the Solar Dynamics Observatory (SDO; \citealt{2013ApJ...779L...7K, 2015ApJ...811L..13W}), and at microwaves with the Nobeyama Radioheliograph simultaneously with the Ramaty High Energy Solar Spectroscopic Imager \citep{2011A&A...525A.112R} and with AIA \citep{2012ApJ...756L..36K}. \citet{2019ApJ...874L...1N} demonstrated that periods of the SUMER oscillations observed in hotter loops which have higher local sound speed, are systematically shorter.   

\citet{2002ApJ...580L..85O} performed the first simulation modeling a SUMER oscillation as a damped slow magnetoacoustic oscillations excited impulsively in a coronal loop. The simulation was made in terms of one-dimensional (1D) MHD, i.e., in the infinite magnetic field approximation. In this approximation it is assumed that the plasma beta is so small that one can neglect displacements of the plasma across the field. In other words, SUMER oscillations were considered to be acoustic oscillations, or, more precisely, as slow magnetoacoustic oscillations with the wave fronts perpendicular to the field. This simple modelling successfully reproduced the observed SUMER oscillation periods (from about 11 to 31 minutes). The oscillation period of the fundamental harmonic was found to be consistent with a simple estimation, $P \approx 2 L/C_\mathrm{s}$, where $L$ is the distance along the loop between the footpoints, and $C_\mathrm{s}$ is the sound speed taken to be uniform along the loop. The rapid damping of the oscillation, on a timescale comparable to observations (from about 6 to 29 minutes), was linked with the high thermal conduction along the field. This pioneering study gave rise to a series of theoretical works which considered different aspects of the wave dynamics. \citet{2004A&A...414L..25N} modelled SUMER oscillations a second parallel harmonic of standing acoustic oscillations, excited by an impulsive energy release in the oscillating loop near its top. More comprehensive numerical modelling of this process, accounting for the effects of gravitational stratification, thermal conduction, optically thin radiative losses, and compression viscosity has been performed by \citet{2019ApJ...884..131R}. Fundamental parallel harmonics could also be readily excited if the energy release is situated at one of the footpoints \citep[see, e.g.,][]{2005A&A...438..713T}. Results of \citet{2005A&A...436..701S} demonstrated that an impulsive energy release induces the fundamental and second parallel harmonics, depending on the location of this pulse along the loop. Generally, the oscillation period of the $n$-th parallel harmonic is $P \approx 2 L/nC_\mathrm{s}$. The oscillation period is weakly sensitive to the inclusion of various dissipative processes in the model, and gravitational stratification. Comparing the phase relation between the Doppler velocity and intensity oscillations, available in some observations, \citet{2003A&A...406.1105W} found it to be about a quarter-period, which is consistent with theoretical modelling \citep{2002ApJ...580L..85O}. Thus, it is commonly accepted that SUMER oscillations are standing slow magnetoacoustic oscillations \citep[see][for a recent comprehensive review]{2021SSRv..217...34W}. 

There is also increasing evidence of SUMER-like QPPs in stellar flares. These have been observed at various wavelengths associated with the thermal emission: X-ray \citep[e.g.,][]{2005A&A...436.1041M, 2006AstL...32..569S, 2013ApJ...778L..28S, 2016A&A...590A...7L, 2016ApJ...830..110C}, ultraviolet \citep[e.g.,][]{2006A&A...458..921W}, and microwave \citep[e.g.,][]{2004AstL...30..319Z}. For example, \citet{2005A&A...436.1041M} found an oscillation with a period of 750 s and damping time of 2000 s in the soft X-ray (SXR) of the active M-type dwarf AT Mic. \citet{2013ApJ...778L..28S} detected oscillations with two statistically significant periodicities, 1261 s and 687 s, in the SXR emission from a flare on the Proxima Centauri. \citet{2016A&A...590A...7L} systematically studied oscillations showing a period of 1000 s in the X-ray of classical T Tauri stars.
Moreover, similar oscillatory patterns, i.e., rapidly-decaying long-period almost-harmonic oscillatory patterns are often detected in the decay phase of stellar flares observed in the white light \citep[e.g.,][]{2013ApJ...773..156A, 2015MNRAS.450..956B, 2015ApJ...813L...5P, 2016MNRAS.459.3659P, 2020A&A...636A..96M}.
For example, \citet{2013ApJ...773..156A} studied oscillations approximated by an exponentially decaying harmonic function with the period of 32 min and the decay time of 46 min in the white light of the dM4.5e star YZ CMi.  \citet{2020A&A...636A..96M} observed white light QPPs superposed of two intrinsic modes with periods of 49 min and 86 min of a young active solar-type star KIC 8414845. \citet{2019ApJ...884..160V} detected quasi-periodic patterns in two stellar flares observed in the white light on M5.5 dwarf Proxima Centauri. The QPP period of the first flare is about 390 min and the QPP of the second flare was found to have two periods, 162 and 324 min. However, it is still not clear whether the white light radiation observed in those flares is of thermal or non-thermal nature.  

There have been several statistical studies dedicated to revealing properties of stellar QPPs. \citet{2016MNRAS.459.3659P} found that 56 flares out of 1439 flares have QPP-like signatures in light curves for stellar flares observed in the white light. The oscillation period was found to correlate with the decay time of QPPs. \citet{2016ApJ...830..110C} analyzed SUMER-like QPPs in solar and stellar flares observed in SXRs. They showed that the power-law slopes between the oscillation periods and decay times of solar and stellar QPPs in solar and stellar flares are similar to each other. \citet{2020ARA&A..58..441N} have compared the power-law slopes of the parameters of SUMER-like QPPs and the SUMER oscillations observed in the solar corona, and found out that they are almost identical to each other. This similarity supports the assumption of a stellar flare occurring in a single loop \citep[e.g.,][]{2014LRSP...11....4R}. 
With the revealed similarity in observations, we can infer that SUMER-like QPP in stellar flares, at least in the observational bands which is clearly associated with the thermal emission, have the same nature as solar SUMER oscillations, i.e., caused by standing slow magnetoacoustic oscillations in stellar coronal loops. In stellar loops, slow magnetoacoustic oscillations have been numerically modelled by \citet{2004A&A...414L..25N} and \citet{2019ApJ...884..131R}, assuming that the major radius (the radius extending from the center of the loop to the axis of the loop, cf. a torus) of the loop is much smaller than the radius of the star. Another equilibrium model was considered by \citet{2018ApJ...856...51R}, who considered acoustic oscillations in a magnetic tube that is several stellar radii long, which possibly links the stellar surface with an accretion disk. 

In this study, we model slow magnetoacoustic oscillations in stellar coronal loops. The aim of this study is to investigate the effect of the stellar loop's size on the period of slow oscillations and validate the applicability of simple estimating relations which link the oscillation period with the length of the oscillating loop. Here, we employ the infinite magnetic field approximation, considering a slow magnetoacoustic wave as a 1D acoustic wave propagating strictly along the magnetic field. We solve a boundary-value problem for 1D acoustic wave equation, varying the ratio of the major loop radius to the radius of the star and accounting for the gravitational stratification.
The paper is organized as follows. In Section \ref{sec:model}, we present the model. Section \ref{sec:results} gives the results of the calculations. Discussion and conclusions are provided in Section \ref{sec:conclusions}.


\section{Model and Governing Equations} \label{sec:model}

\subsection{Description of the model} \label{subsec:models}

\begin{figure*}\centering
\includegraphics[width=0.5\textwidth]{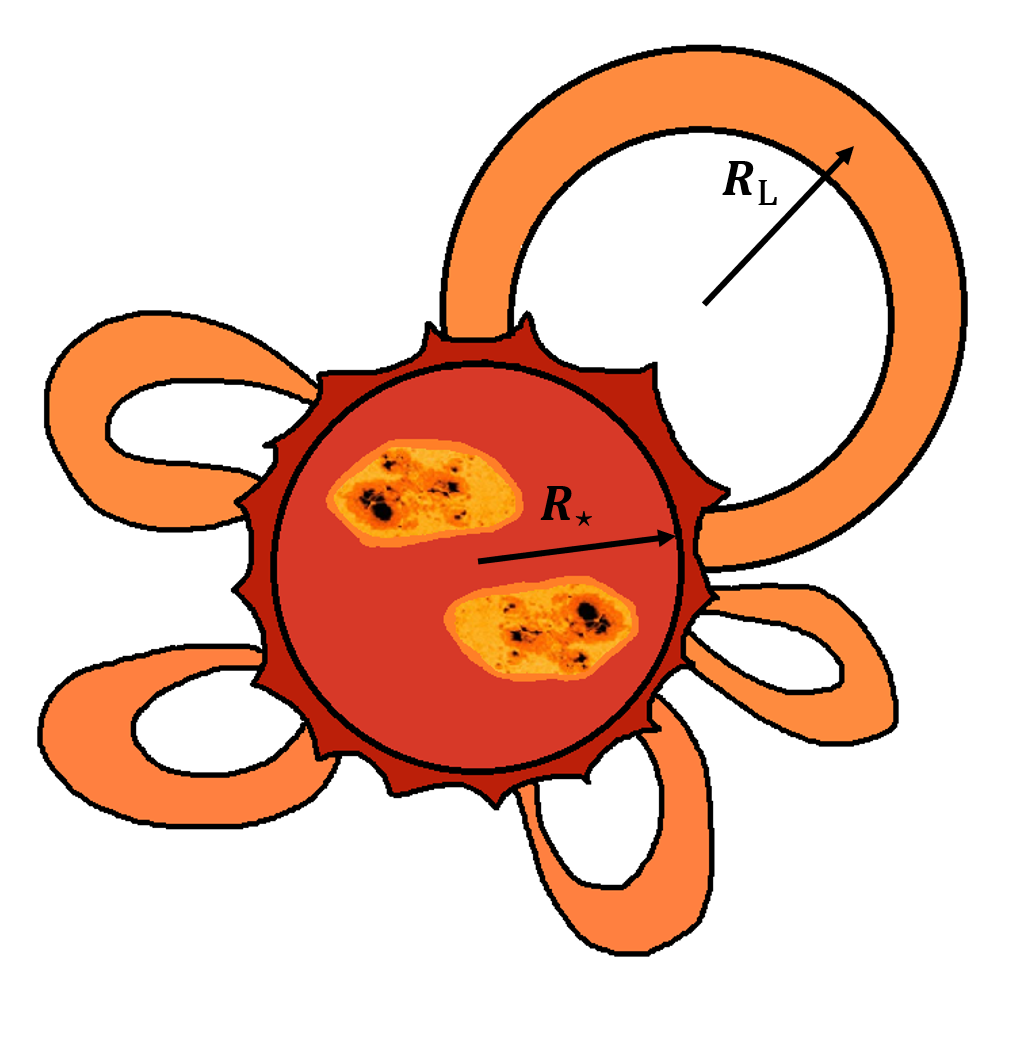}
\includegraphics[width=0.4\textwidth]{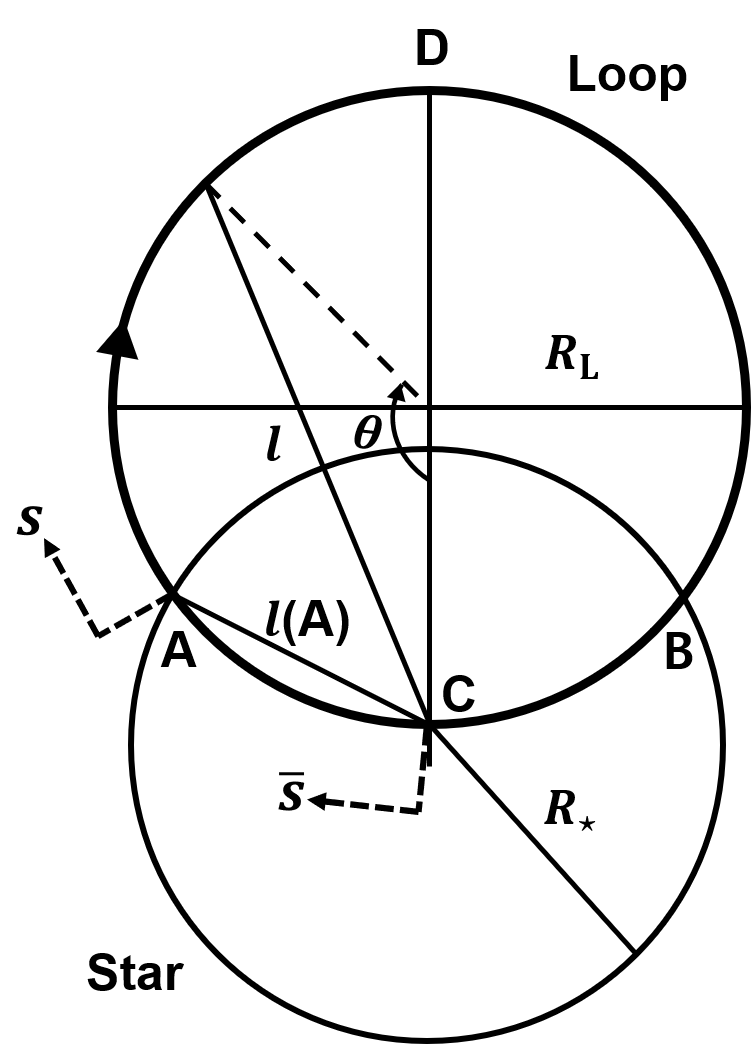}\\ 
\caption{Left: the artistic sketch of a stellar coronal loop. Right: the sketch of the model considered. A variable $\overline{s}$ goes clockwise from point C along the loop and $s$ goes along the loop but from the left footpoint A}.
\label{fig:sketch}
\end{figure*}

In this study we consider oscillatory dynamics of a plasma in a loop in a flaring star such as a red dwarf (M-type star; \citealt{1991ApJ...378..725H, 2013ApJS..207...15K}). Red dwarfs have much stronger atmospheric magnetic fields than the Sun \citep{2012LRSP....9....1R}. Although direct imaging observations of stars other than the Sun are difficult, radio observations of dMe stars have provided several evidences for large coronal loops with height sizes up to several times the stellar radius \citep{1997A&A...317..707A, 2000A&A...353..569P}, such as UX Ari \citep{1999A&A...341..595F}, and Algol \citep{2010Natur.463..207P}. An artistic sketch of a stellar coronal loop modelled in our study is shown in Figure \ref{fig:sketch} (left panel). The corresponding sketch of the model geometry is shown in the right panel. 
We consider a loop of a circular shape. For mathematical convenience we assume that the circle passes through the center of the star (point C).  
The major radius ($R_{\text{L}}$) of the stellar loop is comparable to or bigger than the radius ($R_{\star}$) of the star. The loop is anchored at footpoints A and B. This structure is different from the case of a solar coronal loop, where the major radius of the loop is usually much smaller than the radius of the Sun. The main implications of this difference are the change of the loop from a semi-circle to almost a circle, and the need to account for the dependence of the gravitational acceleration on height. In addition, for a long loop, it is not necessary to account for the effect of the inclination of the loop's plane from the horizon, which may be important for shorter loops \citep[e.g.,][]{2001A&A...379.1106T}. 
The loop's cross-section is constant, and does not change with height. This assumption is consistent with the results obtained for solar coronal loops in the SXR band with the Soft X-ray Telescope/Yohkoh \citep{1992PASJ...44L.181K}, and in the EUV band with the Transition Region and Coronal Explorer (TRACE; \citealt{2000SoPh..193...77W}) and the AIA/SDO \citep{2012A&A...548A...1P}. The situation may be different in stellar loops, but as far as it has not been demonstrated observationally, we can use the solar result. 

The distance from an arbitrary point on the loop to the center of the star, $l$, is described as follows
\begin{equation}\label{eq:chord}
l = \text{chord}(\theta)=2R_{L}\sin \Big( {\frac{\overline{s}}{2R_{L}}} \Big),
\end{equation}
where $\overline{s}$ is the length of an arc from the center of the star, which is $R_{\text{L}}\theta$  increasing in the clockwise direction in the sketch in the right panel of Figure~\ref{fig:sketch}. The value $l$ can be zero at the center of the star when $\overline{s}$ is zero or $2\pi R_{\text{L}}$, and $l$ can be $2R_{\text{L}}$ at the apex of the loop (at D) when $\overline{s}$ is $\pi R_{\text{L}}$. At the left footpoint in the sketch, the value $l(\text{A})$ is approximately $R_{\star}$. Using the relationship in Equation~(\ref{eq:chord}), the values of $\overline{s}$ at the footpoints are $\overline{s}\text{(A)} = 2R_{\text{L}}\text{sin}^{-1}(R_{\star}/{2R_{\text{L}})}$ at the left footpoint A and 
$\overline{s}\text{(B)} = 2\pi R_{\text{L}}-2R_{\text{L}}\text{sin}^{-1}(R_{\star}/{2R_{\text{L}}})$ at the right footpoint B, respectively. These values depend on the ratio of the star radius and the loop major radius.

We focus our attention to oscillatory plasma flows in the direction along the loop axis, i.e., along the local magnetic field, i.e., we employ the infinite field approximation which is justified in the low-$\beta$ limit \citep[see][for discussion]{2021A&A...646A.155D}. The magnetic twist and a plasma profile across the field are neglected in our model. Neglecting the effect of the magnetic field curvature which would lead, in particular to a non-zero centrifugal force which is nonlinear and hence absent from the linear description, we consider the loop as a straight magnetic field line with the density and gravitational acceleration varying along it according to the geometry. In this study, we are interested in the coronal acoustic resonator which is formed in the coronal part of the loop, between footpoints A and B. We need to stress that in the consideration of an acoustic resonator that exists in the coronal part of the loop, the shape of the loop in the sub-surface region is not relevant at all. In particular, under the surface the field lines could have any shape, and not necessarily pass through the centre of the star.

\begin{figure*}\centering
\includegraphics[width=1.0\textwidth]{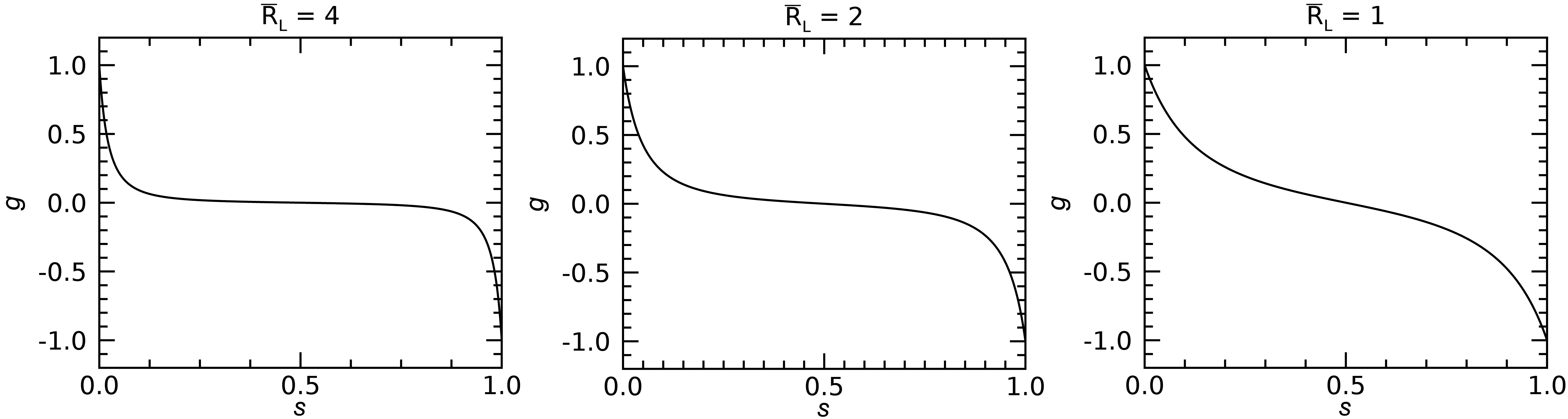}\\ 
\caption{The projection of the gravitational acceleration vector on the loop axis ($s$-axis). The acceleration is normalized by the value of the gravitational acceleration at the footpoint, e.g., $g_A$. The ratio of the loop radius to the radius of the star, $\overline{R}_{\text{L}}$, is 4 (left), 2 (middle), and 1 (right), respectively. $s(0)$ and $s(1)$ are at the left and right footpoints.
\label{fig:g}}
\end{figure*}

In the description of the resonator it is convenient to introduce another variable, $s$, instead of $\overline{s}$. The variable $s$ ranges from $\overline{s}(\text{A})$ to $\overline{s}(\text{B})$. The direction of $s$ is from footpoint A to footpoint B along the coronal part of the loop, with $s=0$ at A.  In the following, the distance $s$ is measured in units of $R_\star$. The projection of the gravitational acceleration on the $s$-axis, $g(s)$, is 
\begin{equation}\label{eq:g}
\displaystyle
g(s) = \frac{GM_\star}{4R^2_{\text{L}}\sin^2(\frac{s+2R_{\text{L}}\text{sin}^{-1}(R_{\star}/{2R_{\text{L}})}}{2R_{\text{L}}})}\cos\Big(\frac{s+2R_{\text{L}}\text{sin}^{-1}(R_{\star}/{2R_{\text{L}})}}{2R_{\text{L}}}\Big),
\end{equation}
where $G$ is the gravitational constant and $M_\star$ is the mass of the star. Hereafter, we use the dimensionless parameter $\overline{R}_{\text{L}}$ which is the ratio of the loop's major radius to the radius of the star. In the following, we consider that $\overline{R}_{\text{L}}=1, 2$, and 4. The profiles of the gravitational acceleration along the loop for three cases of $\overline{R}_{\text{L}}$ are shown in Figure \ref{fig:g}. Here, and throughout the paper the value $s$ shown in figures is normalized so that the footpoints of both loop legs range from 0 to 1 in each case. The gravitational acceleration along the loop, i.e., the projection of the acceleration vector on the direction of the local magnetic field, is maximum at the surface of the star, i.e., at the footpoints, decreases toward the apex of the loop, and becomes zero at the apex where the gravitational acceleration vector is perpendicular to the magnetic field. Figure \ref{fig:g} shows that shorter loops have smoother variation of the projected gravitational acceleration, while in longer loops the gravity is almost constant in a significant part of the loop around the apex. 

\subsection{Equilibrium density profile} \label{subsec:equilibrium}
The temperature of flaring plasmas on dMe stars generally ranges from about 10 MK to 40 MK \citep{1995ApJ...451L..79F, 2004A&ARv..12...71G, 2012ApJ...754..107T}. In the corona, the thermal conduction along the magnetic field is much higher than the thermal conduction in the perpendicular direction. The thermal conductivity is proportional to temperature, thus the higher temperature can reach a uniform state easily. Observed loops in TRACE appeared to be mostly isothermal \citep{1999ApJ...517L.155L, 2000ApJ...541.1059A, 2014LRSP...11....4R}. 
Thus, we assume the plasma in the loop to be isothermal, considering the cases of the temperature $T_{0}\sim10$~MK and $30$~MK. These temperatures correspond to the thermal emission in the EUV and soft X-ray bands.
The sound speed is hence constant too, and is obtained from the ideal gas law, 
\begin{equation}\label{eq:idealgas}
T_{0} = \frac{\overline{\mu}p_{0}}{R\rho_{0}},
\end{equation}
\begin{equation}\label{eq:soundspeed}
C_{s0} = \sqrt{\frac{\gamma R T_{0}}{\overline{\mu}}},
\end{equation}
where $R$ is the gas constant, $\gamma=5/3$, which is the adiabatic index, $\overline{\mu}$ is the mean molecular weight, $p_{0}$ and $\rho_{0}$ are the background pressure and density. 

The mean molecular weight $\overline{\mu}$ is determined by the star's metallicity given by $(2X+3/4Y+1/2Z)^{-1}$, where $X, Y$, and $Z$ is the mass fraction of hydrogen, helium, and the other atoms besides hydrogen and helium, satisfying $X+Y+Z=1$ \citep[e.g.,][]{2016LRSP...13....1A}. For the Sun, $X=0.74, Y=0.25$, and $Z=0.01$, thus $\overline{\mu}$ is about 0.6 \citep{2009ARA&A..47..481A}. \citet{2012ApJ...748...93R} presented the metallicity of 0.28 dex for AD Leonis (a red dwarf star). From this, we assume $Z=0.02$ in this study. \citet{2007MNRAS.382.1516C} showed observed result that $Y$ ranges from about 0.18 to 0.23 when $Z\sim0.02$ for K dwarf. According to this result, we assume $Y=0.23$, then $X$ becomes 0.75. Using these composition, $\overline{\mu}$ for the star we consider is 0.6. Thus, the sound speeds we consider are about 480~km~$\text{s}^{-1}$ for 10~MK and 832~km~$\text{s}^{-1}$ for 30~MK.

\begin{figure*}\centering
\includegraphics[width=1.0\textwidth]{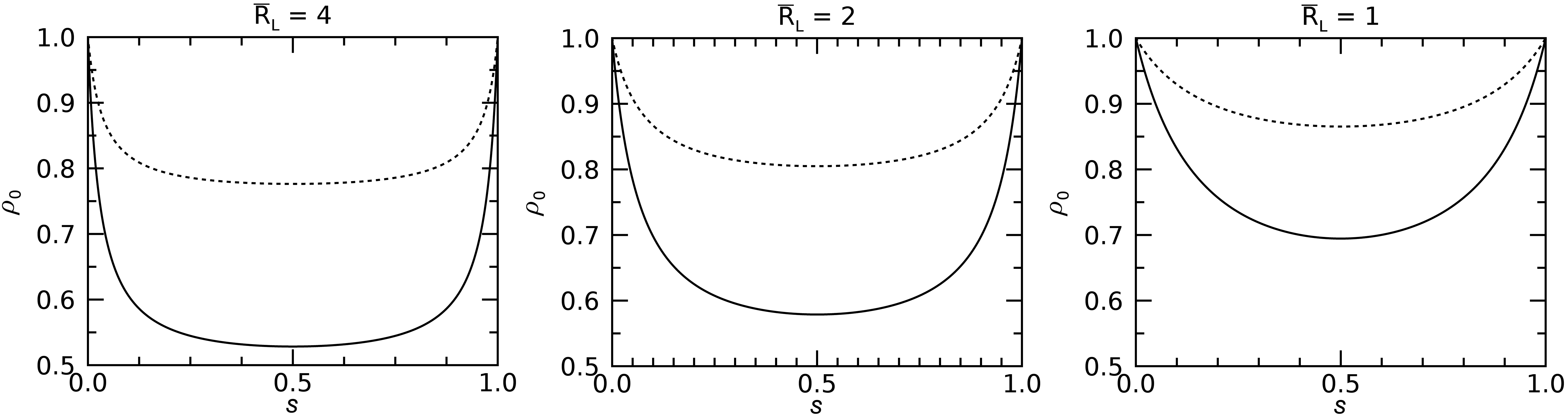}\\ 
\includegraphics[width=1.0\textwidth]{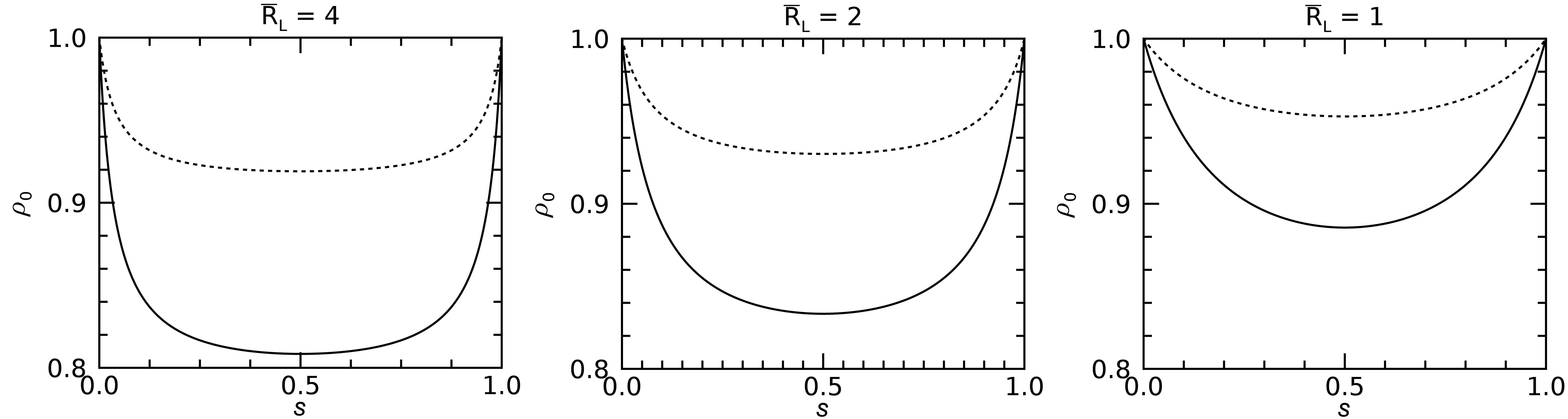}\\
\caption{Equilibrium density $\rho_{0}$  along a stellar loop. The density is normalized by the density value at the footpoint. The equilibrium  temperature ($T_{0}$) is 10~MK (top panel) and 30~MK (bottom panel). The dashed line corresponds to the lighter star and the solid line to the heavier star. The panels are the same as in Figure \ref{fig:g}.
\label{fig:equilibriumdensity}}
\end{figure*}

The hydrostatic equilibrium along the loop, i.e., in the $s$-direction is achieved if
\begin{equation}\label{eq:equilibrium}
\frac{dp_{0}}{ds} = -g\rho_{0}.
\end{equation}
The magnetic field is not included in the condition, as it is always locally parallel to the loop. 
Using Equation~(\ref{eq:equilibrium}) and the uniform equilibrium temperature, we can derive the density profile along the loop as follows,
\begin{equation}\label{eq:density}
\rho_{0}(s) = \rho_{0\text{A}}\exp{\left[\frac{1}{H_{\rho}\sqrt{1-(\frac{R_\star}{2R_{\text{L}}})^2}}\left(\frac{R_\star}{2R_{\text{L}}\sin(\frac{s+2R_{\text{L}}\text{sin}^{-1}(R_{\star}/{2R_{\text{L}})}}{2R_{\text{L}}})}-1 \right)\right]},
\end{equation}
where $H_{\rho} = C_{s0}^2(\gamma g_\text{A}R_\star)^{-1}$ is the local density scale height normalized by the radius of the star and $\rho_{0\text{A}}$ is the density at the left footpoint. The same density is at the other footpoint too. The derivation of the density profile is detailed in Appendix. In the following, we consider two types of stars, a relatively light ($M_\star=0.075M_{\sun}$ and $R_\star=0.36R_{\sun}$) and heavy ($M_\star=0.3M_{\sun}$ and $R_\star=0.57R_{\sun}$) ones. Both considered stars are assumed to have the same density. The density scale height, $H_{\rho}$, and the height of the loop apex above the stellar surface, $H_{L}$, for each case are listed in Table \ref{tab:length}. Figure~\ref{fig:equilibriumdensity} shows the equilibrium density profiles along the loop, normalized by its value at a footpoint. The stratification of the density deepens for the heavier star, longer loop length, and relatively low background temperature. 

\begin{table}[]
\centering
\caption{The density scale height ($H_{\rho}$) and height of the loop apex ($H_{L}$) for different stellar masses, radii and coronal temperatures.}
\label{tab:length}
\begin{tabular}{cccccccc}
\hline
\multirow{2}{*}{$T_{0}$/($10^6$~K)\ \ }  & \multirow{2}{*}{Mass type} & \multicolumn{2}{c}{$\overline{R}_{\text{L}} = 4$} & \multicolumn{2}{c}{$\overline{R}_{\text{L}} = 2$} & \multicolumn{2}{c}{$\overline{R}_{\text{L}} = 1$} \\ \cline{3-8} 
                    &                            & $H_{\rho}$/m          & $H_{L}$/m          & $H_{\rho}$/m          & $H_{L}$/m          & $H_{\rho}$/m          & $H_{L}$/m          \\ \hline
\multirow{2}{*}{10} & Lighter Star               & $8.7\times10^8$      & $1.8\times10^9$    & $8.9\times10^8$      & $7.5\times10^8$      & $1.0\times10^9$      & $2.5\times10^8$     \\ \cline{2-8} 
                    & Heavier Star               & $5.5\times10^8$      & $2.8\times10^9$      & $5.7\times10^8$      & $1.2\times10^9$      & $6.3\times10^8$      & $4.0\times10^8$      \\ \hline
\multirow{2}{*}{30} & Lighter Star               & $2.6\times10^9$    & $1.8\times10^9$      & $2.7\times10^9$     & $7.5\times10^8$     & $3.0\times10^9$      & $2.5\times10^8$      \\ \cline{2-8} 
                    & Heavier Star               & $1.7\times10^9$      & $2.8\times10^9$    & $1.7\times10^9$      & $1.2\times10^9$      & $1.9\times10^9$      & $4.0\times10^8$      \\ \hline
\end{tabular}
\end{table}

\subsection{Governing equations} \label{subsec:equations}

We consider a standing slow magnetoacoustic oscillation along the loop axis ($s$-direction). The plasma perturbations are described by the infinite field approximation equations, i.e., equations of 1D acoustics, 
\begin{equation}\label{eq:continuity}
\frac{\partial \rho}{\partial t} = -\frac{\partial}{\partial s}(\rho_{0}V),
\end{equation}
\begin{equation}\label{eq:motion}
\rho_{0}\frac{\partial V}{\partial t} = -\frac{\partial p}{\partial s}-g\rho,
\end{equation}
\begin{equation}\label{eq:energy}
\frac{\partial p}{\partial t} = -v\frac{\partial p_{0}}{\partial s}-\gamma p_{0}\frac{\partial V}{\partial s},
\end{equation}
where $\rho$, $V$, and $p$ are perturbations of density, velocity, and pressure. We neglect the heating and cooling terms in the energy equation because in this study we are only interested in the resonant periods. Equations~(\ref{eq:continuity})--(\ref{eq:energy}) can be combined into one equation,
\begin{equation}\label{eq:ode}
\frac{\partial^2 V}{\partial s^2} = - \left( \frac{\gamma+1}{\gamma p_{0}}\frac{\text{d} p_{0}}{\text{d}s}+\frac{g}{C_{s0}^2} \right)\frac{\partial V}{\partial s}-\left(\frac{1}{\gamma p_{0}}\frac{\text{d}^2 p_{0}}{\text{d}s^2}+\frac{g}{\gamma p_{0}}\frac{\text{d} \rho_{0}}{\text{d}s}\right)V+\frac{1}{C_{s0}^2}\frac{\partial^2 V}{\partial t^2}.
\end{equation}
In the following, we assume a harmonic dependence on time, i.e., the perturbations on the time $t$ proportional to $\exp(i\omega t)$, where $\omega$ is the angular frequency. Using the expression (\ref{eq:equilibrium}) and this assumption, Equation~(\ref{eq:ode}) can be simplified as
\begin{equation}\label{eq:ode_simplified}
\frac{\text{d}^2 V}{\text{d} s^2} = \frac{\gamma g}{C_{s0}^2}\frac{\text{d} V}{\text{d} s}+\left(\frac{1}{C_{s0}^2}\frac{\text{d} g}{\text{d} s}-\kappa^2\right)V,
\end{equation}
where $\kappa \equiv \omega/C_{s0}$. When $g$ is independent of the distance from the star, i.e., in the solar case when loops are much shorter than the solar radius, Equation~(\ref{eq:ode_simplified}) coincides with Equation~(6) in \citet{2015A&A...582A..57A}.

Using Equation~(\ref{eq:ode_simplified}) supplemented by the boundary conditions in the $s$-direction are $V(s=$A)$=V(s=$B)$=0$, we investigate standing slow magnetoacoustic oscillations in an isothermal loop with the major radius comparable or bigger than the radius of the star, accounting for the effects of density stratification, background temperature, the mass of the star, and the ratio of the major radius of the loop to the star radius. We solve a boundary-value problem constituted by Equation (\ref{eq:ode_simplified}) with the boundary conditions for eigenvalues (i.e., the discrete values of $\kappa$) and eigen functions ($V(s)$ corresponding to specific eigenvalues) numerically. The numerical domain is from the left footpoint A to right footpoint B, corresponding to a stellar loop (see Figure \ref{fig:sketch}, right). The boundary problem was solved by the standard shooting method. 


\section{Results} \label{sec:results}

\begin{figure*}\centering
\includegraphics[width=1.0\textwidth]{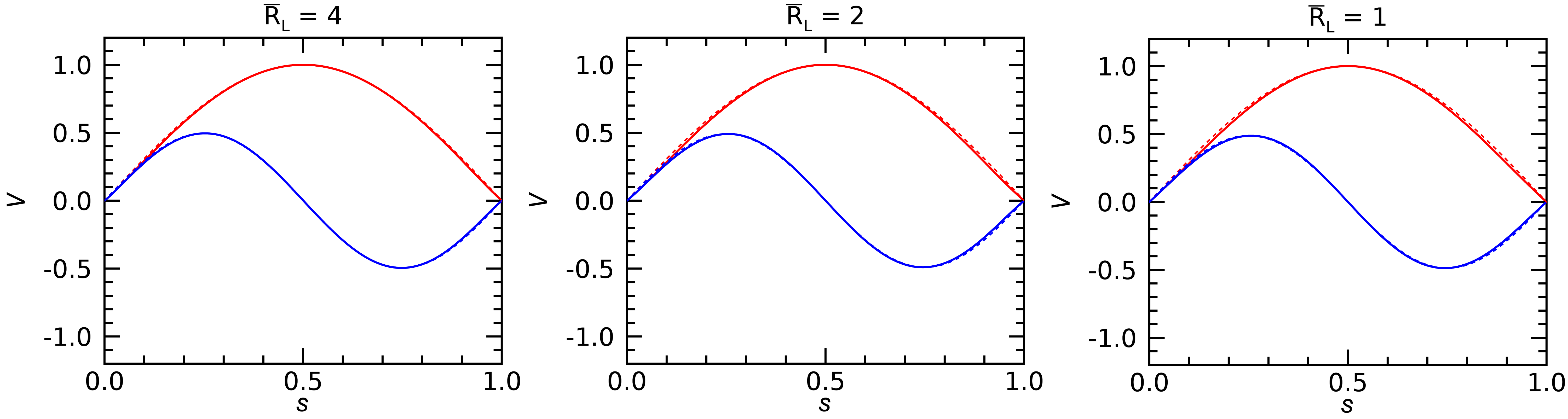}\\
\caption{Velocity perturbations along a stellar loop with different major radii. The star mass is $0.3M_{\sun}$. The background temperature $T_{0}$ is 10~MK. Values of $V$ are normalized by $C_{s0}$. The red and blue solid lines correspond to the fundamental and second harmonics, respectively. The dashed lines (practically coinciding with the solid lines) show the same but for a loop with a constant density. The amplitudes are arbitrary. The distance along the loop is normalised to the length of the loop.
\label{fig:velocity}}
\end{figure*}

\begin{figure*}\centering
\includegraphics[width=1.0\textwidth]{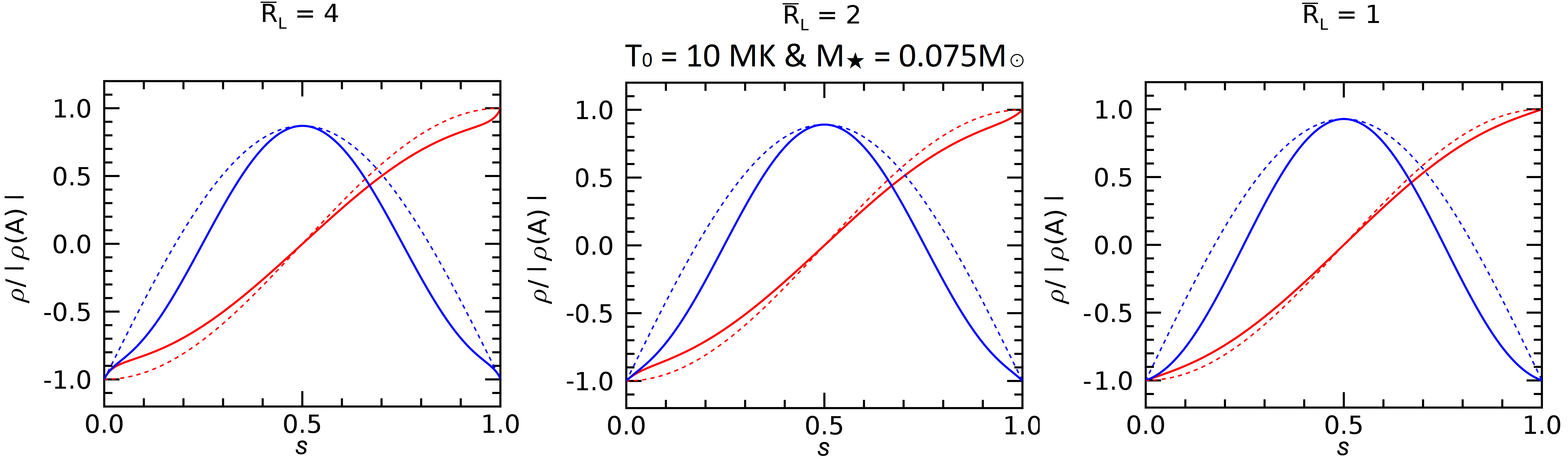}\\ 
\includegraphics[width=1.0\textwidth]{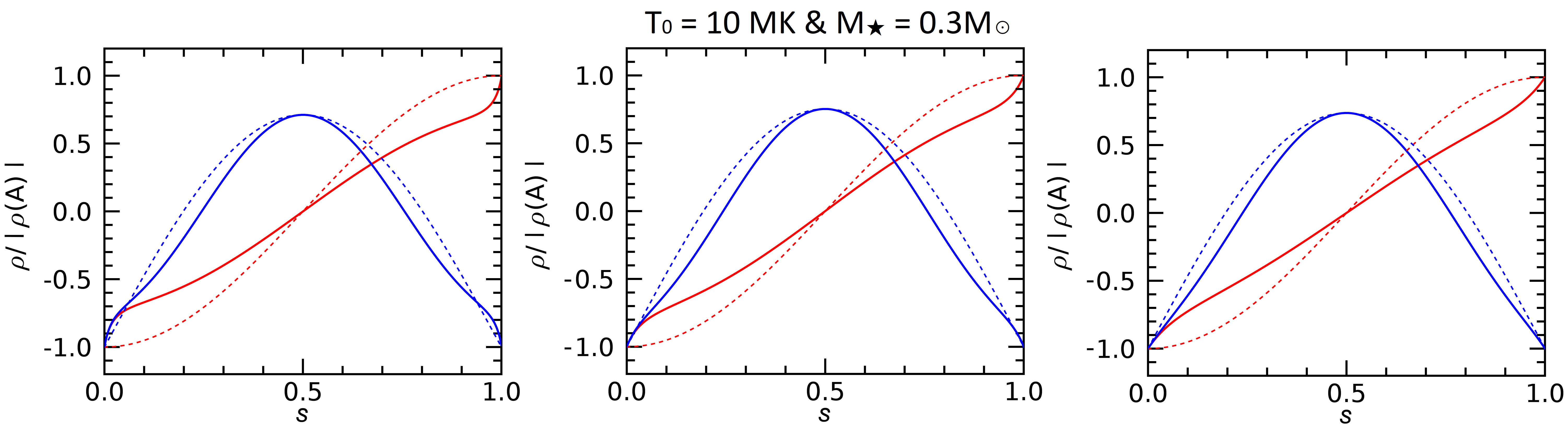}\\
\includegraphics[width=1.0\textwidth]{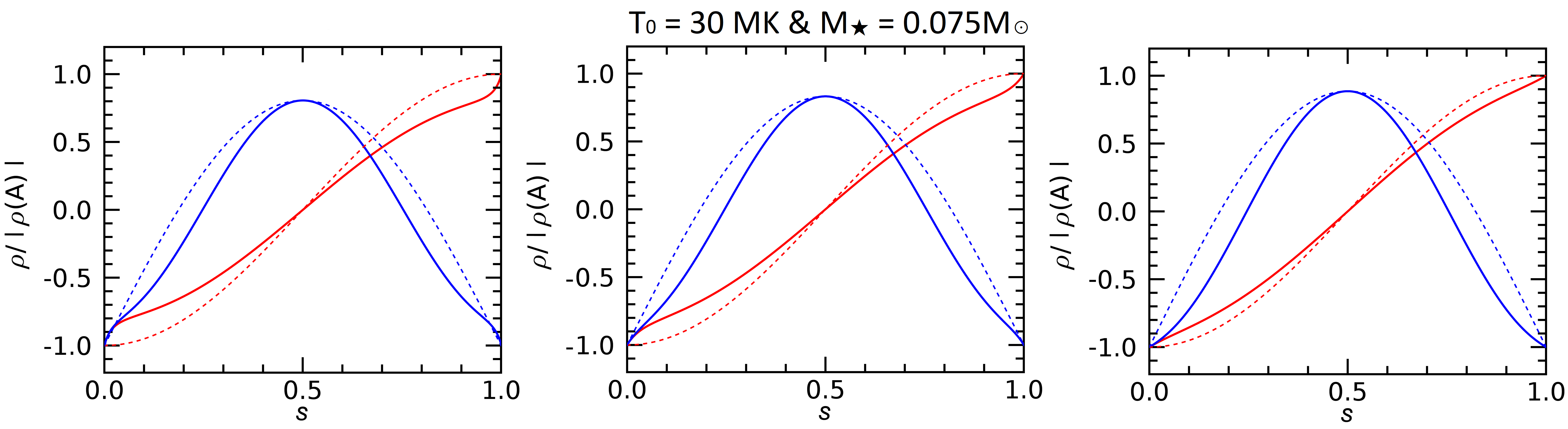}\\ 
\includegraphics[width=1.0\textwidth]{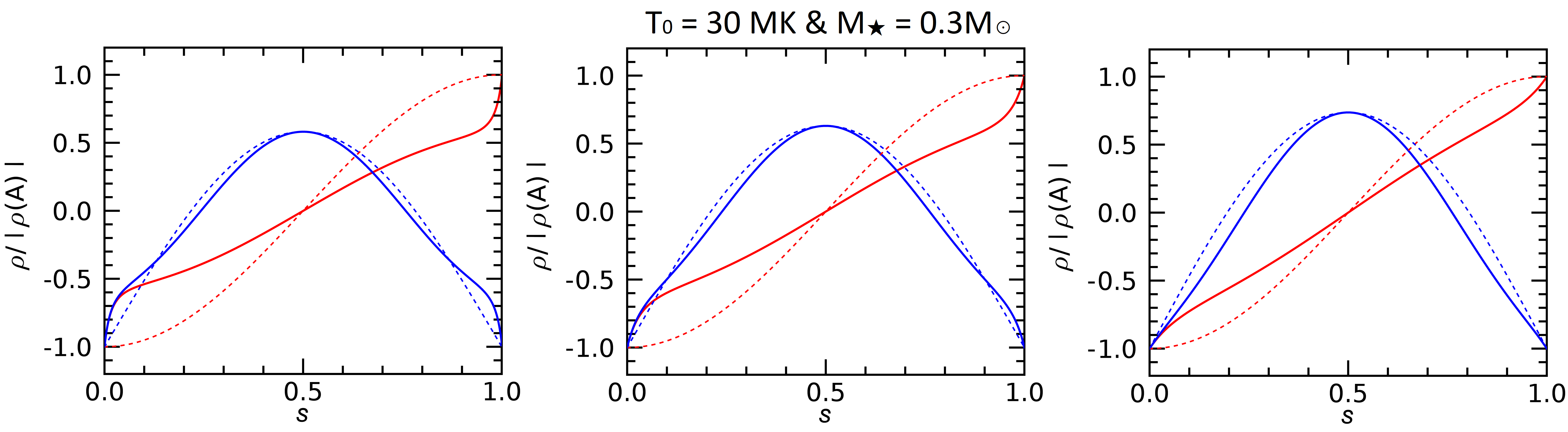}\\
\caption{Density perturbations along a stellar loop for a lighter star (first and third row panel) and a heavier star (second and fourth row panel). The background temperature $T_{0}$ is 10 MK (first and second row panel) and 30 MK (third and fourth row panel). The parameter $\overline{R}_{\text{L}}$ which is the ratio of $R_{L}$ to the radius of the star $R_\star$, is 4 (left), 2 (middle), and 1 (right), respectively. The red and blue solid lines correspond to the fundamental and second harmonics, respectively. The dashed lines show the same but for a loop with a constant density.  
All values are normalized by the density perturbation amplitude at a footpoint. 
\label{fig:normalizeddensity}}
\end{figure*}

Eigen functions corresponding to two lowest parallel harmonics of standing slow magnetoacoustic waves in a stellar loop are exposed in Figures~\ref{fig:velocity} and \ref{fig:normalizeddensity} for the parallel velocity and perturbed density, respectively. As in the solar case \citep[e.g.,][]{2021SSRv..217...34W}, the density and velocity perturbations along the loop are phase-shifted with respect to each other, with the maxima of the velocity oscillations corresponding to the nodes of the density oscillations. At the footpoints, the velocity perturbations are zero, while the density perturbations are maximum.
For comparison, we show eigen functions of a loop without the stratification, i.e., a uniform loop, but of the same size and temperature, and on the same stars. 
In Figure~\ref{fig:velocity} we show the structure of the velocity perturbations only for the case of a heavier star and temperature of 10~MK. For all three major radii of the loop, with the applied normalisation the structure of the oscillations is practically the same. Moreover, structures of the oscillations, obtained for a lighter star and in a hotter loop are practically indistinguishable from the shown dependencies. A very slight difference could be seen for a heavier star, shorter loop length, and lower background temperature. 

\begin{figure*}\centering
\includegraphics[width=1.0\textwidth]{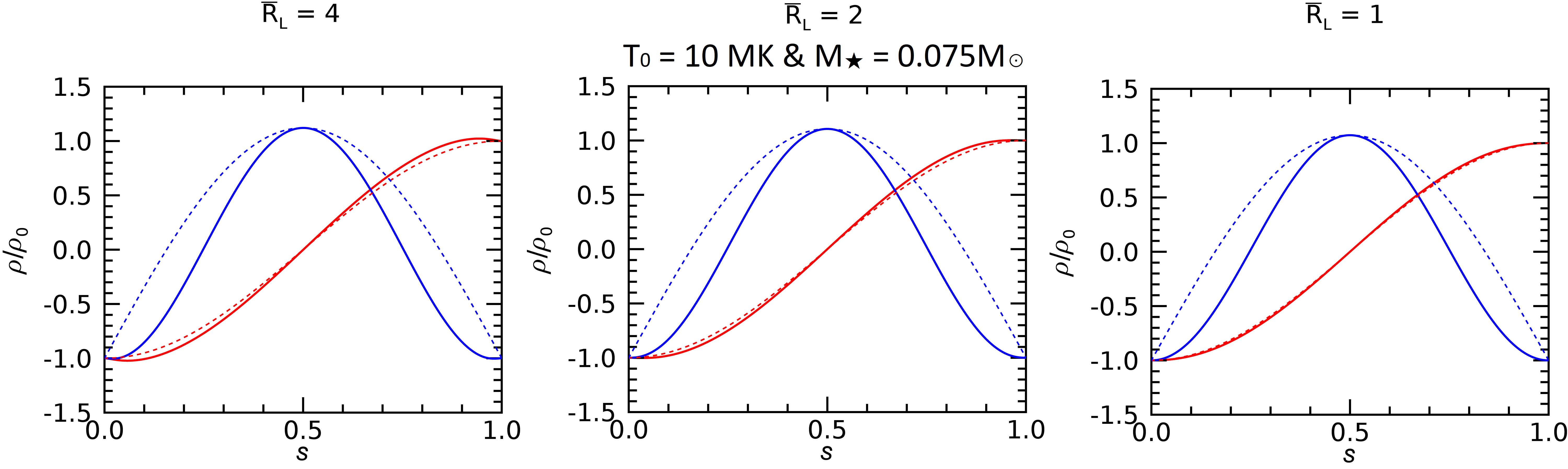}\\
\includegraphics[width=1.0\textwidth]{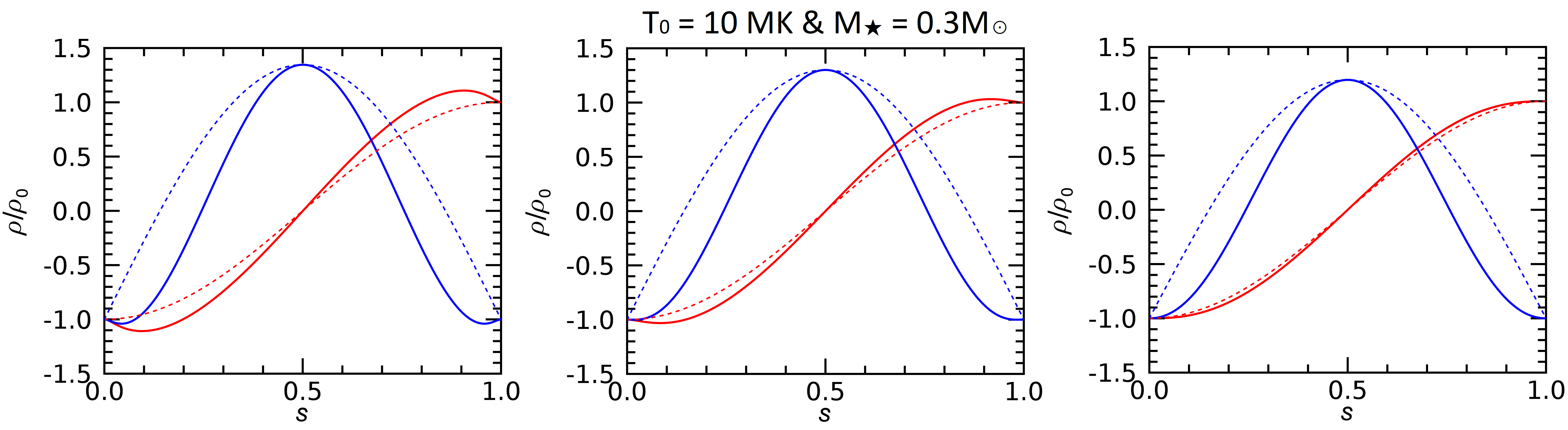}\\
\includegraphics[width=1.0\textwidth]{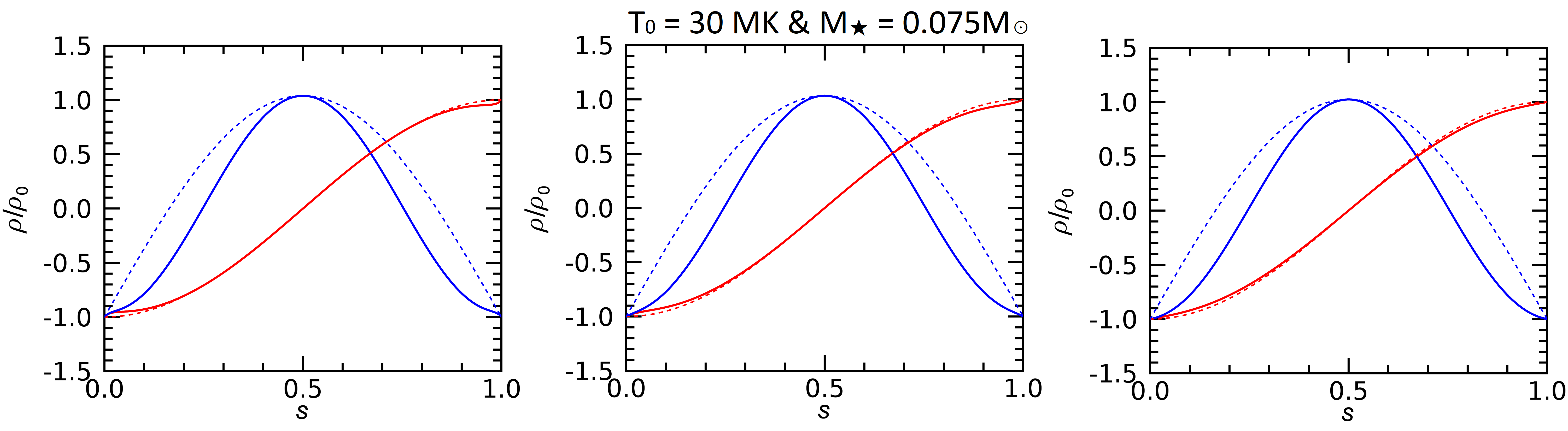}\\
\includegraphics[width=1.0\textwidth]{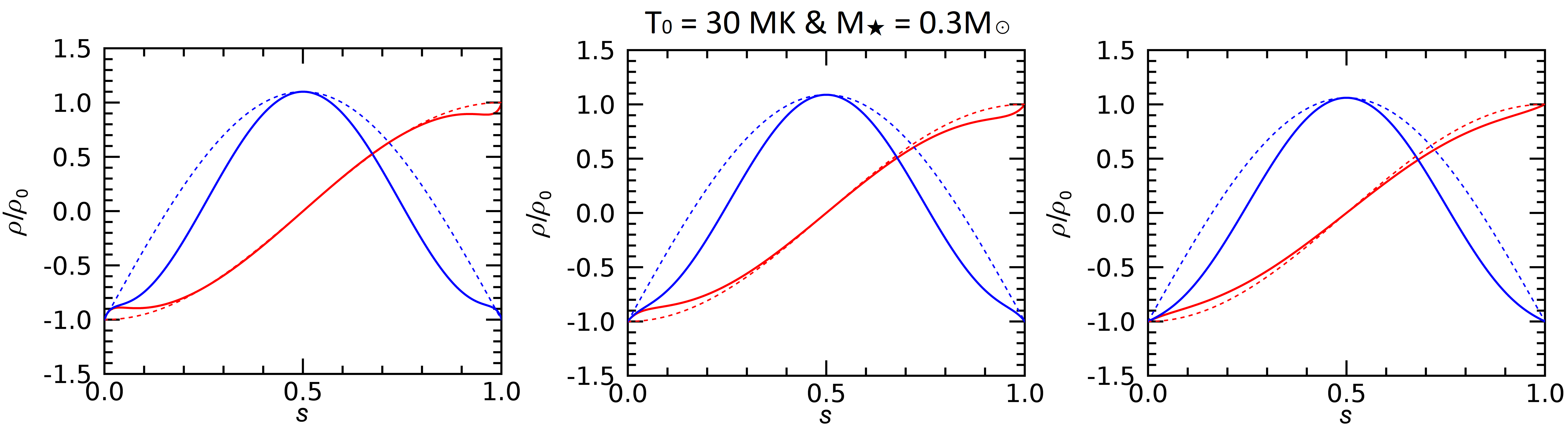}\\
\caption{Relative density perturbations along a stellar loop, which are density perturbations divided by the local equilibrium density. The panels and notations are the same as in Figure \ref{fig:normalizeddensity}.
\label{fig:localizeddensity}}
\end{figure*}

In contrast, the differences between the spatial structures of the density perturbations in loops with different parameters are much more pronounced, especially for heavier stars, longer loops, and higher temperatures (see Figure~\ref{fig:normalizeddensity}).
Figure \ref{fig:localizeddensity} shows the spatial structure of the local density perturbations, divided by the local value of the equilibrium density, i.e., the relative perturbation of the density. The relative density perturbations are enhanced at footpoints for both two harmonics. The relative density perturbations for heavier stars, longer loops and lower temperatures are stronger.

The oscillation period is determined as $2\pi/\omega$ by the numerically determined eigenvalue $\kappa$. The period of the modes in the non-stratified model, used for comparison, is calculated as $2L/(C_{s0}n)$, where $n$ is the parallel harmonic number and $L$ is the length of the loop (the distance between the footpoints $s=$A and $s=$B along the loop, see Figure \ref{fig:sketch}, right). The oscillation periods are summarised in Table \ref{tab:period}. 
We find that standing acoustic oscillations in the model with the stratification have slightly longer oscillation periods than the oscillations in the model with a uniform density. This result is consistent with the result of \citet{2004ApJ...605..493M} who compared the acoustic oscillation periods in a stratified and non-stratified corona in the solar case, i.e., for a constant gravity. Oscillation periods are found to be longer for heavier star, longer loops, and lower temperatures. 

\begin{table}[h]
\centering
\caption{Oscillation periods calculated in the stratified ($P_{\text{c}}$) and uniform ($P_{\text{u}}$) loop models. The periods are measured in minutes. The integer $n$ is the harmonic number.}
\label{tab:period}
\begin{tabular*}{\textwidth}{@{\extracolsep{\fill}}cccccccccccc}
\hline
\multirow{2}{*}{$T_{0}$ (MK)} & \multirow{2}{*}{Mass type} & \multirow{2}{*}{$n$} & \multicolumn{3}{c}{$\overline{R}_{\text{L}} = 4$} & \multicolumn{3}{c}{$\overline{R}_{\text{L}} = 2$} & \multicolumn{3}{c}{$\overline{R}_{\text{L}} = 1$} \\ \cline{4-12} 
 &  &  & $\kappa^{'}$ & $P_{\text{c}}$ & $P_{\text{u}}$ & $\kappa^{'}$ & $P_{\text{c}}$ & $P_{\text{u}}$ & $\kappa^{'}$ & $P_{\text{c}}$ & $P_{\text{u}}$ \\ \hline
 &  & 1 & 0.135 & 403.34 & 401.37 & 0.296 & 184.03 & 183.03 & 0.746 & 73.04 & 72.70 \\
 & Lighter Star & 2 & 0.271 & 201.25 & 200.69 & 0.594 & 91.75 & 91.51 & 1.497 & 36.41 & 36.35 \\
\multirow{2}{*}{10} &  & 3 & 0.407 & 134.04 & 133.79 & 0.892 & 61.10 & 61.01 & 2.248 & 24.25 & 24.23 \\ \cline{2-12} 
 &  & 1 & 0.134 & 644.36 & 637.38 & 0.294 & 294.21 & 290.65 & 0.742 & 116.67 & 115.45 \\
 & Heavier Star & 2 & 0.270 & 320.68 & 318.69 & 0.593 & 146.08 & 145.32 & 1.495 & 57.92 & 57.72 \\
 &  & 3 & 0.406 & 213.27 & 212.46 & 0.891 & 97.16 & 96.88 & 2.246 & 38.54 & 38.48 \\ \hline
 &  & 1 & 0.136 & 232.13 & 231.73 & 0.297 & 105.87 & 105.67 & 0.749 & 42.04 & 41.97 \\
 & Lighter Star & 2 & 0.271 & 115.98 & 115.87 & 0.595 & 52.88 & 52.84 & 1.499 & 21.00 & 20.99 \\
\multirow{2}{*}{30} &  & 3 & 0.407 & 77.30 & 77.24 & 0.893 & 35.24 & 35.22 & 2.249 & 14.00 & 13.99 \\ \cline{2-12} 
 &  & 1 & 0.135 & 369.52 & 367.99 & 0.297 & 168.58 & 167.81 & 0.747 & 66.92 & 66.65 \\
 & Heavier Star & 2 & 0.271 & 184.44 & 184.00 & 0.595 & 84.09 & 83.90 & 1.498 & 33.37 & 33.33 \\
 &  & 3 & 0.407 & 122.86 & 122.66 & 0.893 & 56.00 & 55.94 & 2.248 & 22.23 & 22.22 \\ \hline
\end{tabular*}
\tablecomments{$\kappa^{'}$ is normalized $\kappa$ to the inverse of the star's radius.}
\end{table}

\section{Discussion and Conclusions} \label{sec:conclusions}

In this study, we modelled linear standing slow magnetoacoustic oscillations in a stellar coronal loop in the infinite field approximation, i.e., in terms of 1D acoustics. The applicability of this approximation to loops filled in with a low-$\beta$ plasma has been justified by \citet{2021A&A...646A.155D}. We investigated effects of the gravitational stratification, star mass, background temperature, and loop length on properties of the oscillations. The loop in its coronal part is assumed to be isothermal and of a circular shape. The major radius of the loop is taken to be comparable or greater than the radius of the star, which requires accounting for the variation of the gravitational acceleration with height. In equilibrium, the density varies along the loop according to stratification. An ordinary differential equation which describes the spatial structure of the oscillation along the loop is derived. Together with boundary conditions at the loop's footpoints this differential equation constitutes an eigenvalue problem which determines oscillation periods of slow magnetoacoustic modes and spatial structure of the perturbations. In general, these parameters of the oscillations are determined by the value of the temperature, which prescribes the sound speed $C_{s0}$. In addition, sound waves are sensitive to the density profile $\rho_{0}(s)$ which determines the acoustic impedance $\rho_{0}(s)C_{s0}$. Our attention was paid to the lowest parallel harmonics only, which is motivated by spatially-resolving observations of oscillations of this kind in solar coronal loops \citep[see, e.g.,][]{2019ApJ...874L...1N, 2021SSRv..217...34W}. This model could also be applied to the mechanisms for quasi-periodic triggering of magnetic reconnection by slow magnetoacoustic waves progressing along a neutral line in a coronal arcade \citep{2011ApJ...730L..27N}. 

An important feature of the considered model is that with the increase in the major radius of the loop with respect to the radius of the star, a significant part of the loop has almost a constant value of the equilibrium density. Together with the constant temperature, and also as the wave parameters are mainly determined by the segment of the loop where the wave amplitude has a maximum, i.e., near the top of the loop for the fundamental harmonic \citep[see the discussion in][]{2009SSRv..149....3A}, it suggests that resonant acoustic periods of such a long loop could be estimated using a simple formula $2L/nC_{s0}$ with a positive integer $n$ corresponding to the parallel harmonic number. This result is confirmed by the solution to the eigenvalue problem, performed in our study. The periods in the stratified loop are slightly longer than in the non-stratified loop. Such a behaviour is consistent with the results obtained for stratified solar coronal loops in a plane atmosphere \citep{2004ApJ...605..493M}.  
The spatial structure of the velocity perturbations in two lowest acoustic modes of the loop is almost indistinguishable from the case of a loop with constant equilibrium density. In contrast, the spatial structure of density perturbations shows a significant deviation from the uniform case, which is especially pronounced near the footpoints for the fundamental mode, and in the loop's legs for the second harmonic. A similar modification of the eigen functions was found for the spatial structure of the relative density perturbations determined as the ratio of the density perturbations to the local value of the equilibrium density. The difference in the effects of the stratification on the spatial structures of the velocity and density perturbations in the loop can be justified by the following reasoning. In a uniform loop, a relative perturbation of the velocity, i.e., divided by the sound speed, is proportional to the relative perturbation of the density, i.e., the density perturbation divided by the equilibrium density. Thus, in the considered case of the constant sound speed and varying (especially near footpoints) equilibrium density, one could expect the departure of the spatial structure of the relative density perturbation from the uniform density case to be more pronounced. 

Our findings have important implications in MHD seismology of flaring loops. 
Here we would like to stress that in the context of seismology, a justification of a simpler model or a simpler estimating expression is a very valuable result that allows one to link observables with physical parameters of interest.
In particular, the high precision of the simple estimating expression $P = 2L/nC_{s0}$, established by our study, allows for estimating the stellar loop length by two observables, the oscillation period of the thermal emission modulation and the temperature.
Moreover, in contrast with the solar case, there is no need for the angle between the plane of the loop and the local plane of the solar surface, which, in general, requires spatially-resolving stereoscopic observations. This is because in the case of the plane atmosphere this angle reduces the height of the loop top from the surface, while in the case with the loop much larger than the stellar radius, the angle between the loop’s plane and the local tangent to the stellar surface does not change the distance from the loop top to the surface. 

Consider implications of the results obtained to the QPP observed in stellar flares.
\citet{2013ApJ...778L..28S} observed 21 min and 12 min soft X-ray QPP in a flare at Proxima Centauri. The mass of the star is $0.123 M_\odot$, and the radius is $10^8$~m. Using the estimating expression, we obtain that the fundamental parallel harmonic of standing slow waves has a period in this range for the temperature $2\times 10^7$~K and loop length of $4\times 10^8$~m. This value of the loop length corresponds to the radius of a circular loop of about $6.4\times 10^7$~m, which is about the radius of the host star, justifying the plausibility of the developed theory. 
Similar estimations are applicable to the loops which hosted the flares analysed by \citet{2016ApJ...830..110C}, who reported the detection of SXR QPPs with oscillation periods ranging from about 2 minutes to 70 minutes at several different stars, \citet{2018RSPTA.37670253L} who detected 13-min and 33-min QPP in a flare on the M dwarf AT Mic \citep[see also][]{2005A&A...436.1041M}, and \citet{2009ApJ...697L.153P} who found 17-min QPP at $\xi$~Boo. 
The main modulation of the observed emission should come from the legs of the oscillating loops, where the relative perturbation of the density of the emitting plasma is strongest. 

Another detection of QPP in flares at Proxima Centauri showed periods ranging from 162 min to 390 min, in white light \citep{2019ApJ...884..160V}. White-light QPP could be produced by periodic triggering of magnetic reconnection by slow magnetoacoustic oscillations \citep{2011ApJ...730L..27N}. In this case, the temperature of the loops forming the arcade could be low than in a flaring loop. 
For example, for the oscillating loop of the star, a 200~min periodicity from the expression $P = 2L/C_{s0}$ can be obtained for the temperature of $10^6$~K and the loop length of $10^9$~m, i.e., 10~$R_\star$. Perhaps, such a long loop is not realistic, and these QPP should be associated with another mechanism \citep[see, e.g.][]{2021SSRv..217...66Z}.
On the other hand, white-light QPP with periods of 32-min, detected by \citet{2013ApJ...773..156A}; 78-min and 32-min, detected by \citet{2015ApJ...813L...5P}; and 8--90 min detected by  \citet{2016MNRAS.459.3659P}, are consistent with the slow magnetoacoustic oscillation mechanism. 

The developed model is readily expandable. It creates a basis for more elaborated studies of slow oscillations in stellar loops with non-uniform temperature, i.e., in which thermal conduction and radiative losses are balanced by heating. In particular, it will allow one to investigate effects of damping and excitation of the oscillations, for example, by the effect of thermal misbalance \citep[e.g.,][]{2019PhPl...26h2113Z}. 


\appendix
\section{The equilibrium density profile along the loop}

The equilibrium density profile along the isothermal coronal loop can be derived from the Equation~(\ref{eq:idealgas}) and Equation~(\ref{eq:equilibrium}) as follows.

\begin{equation}\label{eq:density2}
\begin{split}
\rho_{0}(s) & =
\rho_{0\text{A}}\ \exp\left(-\frac{\gamma}{C_{s0}^2}\int_{0}^{s} g(s^{'})\ \text{d}s^{'}\right) \\
& = \rho_{0\text{A}}\ \exp\left(-\frac{\gamma}{C_{s0}^2}\int_{0}^{s} \frac{GM_\star}{4R^2_{\text{L}}\sin^2(\frac{s^{'}+\overline{s}(\text{A})}{2R_{\text{L}}})}\cos\Big(\frac{s^{'}+2R_{\text{L}}\text{sin}^{-1}(R_{\star}/{2R_{\text{L}})}}{2R_{\text{L}}}\Big)\ \text{d}s^{'}\right) \\
& = \rho_{0\text{A}}\ \exp\left[-\frac{\gamma GM_\star}{C_{s0}^2} \left( \left( -\frac{1}{2R_{\text{L}}\sin(\frac{s+2R_{\text{L}}\text{sin}^{-1}(R_{\star}/{2R_{\text{L}})}}{2R_{\text{L}}})} \right) -  \left( -\frac{1}{2R_{\text{L}}\sin(\frac{2R_{\text{L}}\text{sin}^{-1}(R_{\star}/{2R_{\text{L}})}}{2R_{\text{L}}})} \right) \right) \right] \\
& = \rho_{0\text{A}}\ \exp\left[\frac{\gamma GM_\star}{C_{s0}^2}\left(\frac{1}{2R_{\text{L}}\sin(\frac{s+2R_{\text{L}}\text{sin}^{-1}(R_{\star}/{2R_{\text{L}})}}{2R_{\text{L}}})}-\frac{1}{R_\star}\right)\right] \\
& = \rho_{0\text{A}}\exp{\left[\frac{1}{H_{\rho}\sqrt{1-(\frac{R_\star}{2R_{\text{L}}})^2}}\left(\frac{R_\star}{2R_{\text{L}}\sin(\frac{s+2R_{\text{L}}\text{sin}^{-1}(R_{\star}/{2R_{\text{L}})}}{2R_{\text{L}}})}-1 \right)\right]}
\end{split}
\end{equation}


\begin{acknowledgments}

This work was supported by the Basic Science Research Program through the National Research Foundation (NRF) funded by the Ministry of Education (NRF-2021R1I1A1A01040372). 
V.M.N. acknowledges support from the STFC consolidated grant ST/T000252/1.
Y.J.M. acknowledges support from the Korea Astronomy and Space Science Institute under the R\&D program (Project No. 2022-1-850-05) supervised by the Ministry of Science and ICT. The authors are grateful to the referee for valuable comments.

\end{acknowledgments}




\bibliography{StellarLoop_Lim}{}

\begin{thebibliography}{}
\expandafter\ifx\csname natexlab\endcsname\relax\def\natexlab#1{#1}\fi
\providecommand{\url}[1]{\href{#1}{#1}}
\providecommand{\dodoi}[1]{doi:~\href{http://doi.org/#1}{\nolinkurl{#1}}}
\providecommand{\doeprint}[1]{\href{http://ascl.net/#1}{\nolinkurl{http://ascl.net/#1}}}
\providecommand{\doarXiv}[1]{\href{https://arxiv.org/abs/#1}{\nolinkurl{https://arxiv.org/abs/#1}}}

\bibitem[{{Afanasyev} \& {Nakariakov}(2015)}]{2015A&A...582A..57A}
{Afanasyev}, A.~N., \& {Nakariakov}, V.~M. 2015, \aap, 582, A57,
  \dodoi{10.1051/0004-6361/201526530}

\bibitem[{{Alef} {et~al.}(1997){Alef}, {Benz}, \&
  {Guedel}}]{1997A&A...317..707A}
{Alef}, W., {Benz}, A.~O., \& {Guedel}, M. 1997, \aap, 317, 707

\bibitem[{{Allende Prieto}(2016)}]{2016LRSP...13....1A}
{Allende Prieto}, C. 2016, Living Reviews in Solar Physics, 13, 1,
  \dodoi{10.1007/s41116-016-0001-6}

\bibitem[{{Andries} {et~al.}(2009){Andries}, {van Doorsselaere}, {Roberts},
  {Verth}, {Verwichte}, \& {Erd{\'e}lyi}}]{2009SSRv..149....3A}
{Andries}, J., {van Doorsselaere}, T., {Roberts}, B., {et~al.} 2009, \ssr, 149,
  3, \dodoi{10.1007/s11214-009-9561-2}

\bibitem[{{Anfinogentov} {et~al.}(2013){Anfinogentov}, {Nakariakov},
  {Mathioudakis}, {Van Doorsselaere}, \& {Kowalski}}]{2013ApJ...773..156A}
{Anfinogentov}, S., {Nakariakov}, V.~M., {Mathioudakis}, M., {Van
  Doorsselaere}, T., \& {Kowalski}, A.~F. 2013, \apj, 773, 156,
  \dodoi{10.1088/0004-637X/773/2/156}

\bibitem[{{Aschwanden} {et~al.}(2000){Aschwanden}, {Nightingale}, \&
  {Alexander}}]{2000ApJ...541.1059A}
{Aschwanden}, M.~J., {Nightingale}, R.~W., \& {Alexander}, D. 2000, \apj, 541,
  1059, \dodoi{10.1086/309486}

\bibitem[{{Asplund} {et~al.}(2009){Asplund}, {Grevesse}, {Sauval}, \&
  {Scott}}]{2009ARA&A..47..481A}
{Asplund}, M., {Grevesse}, N., {Sauval}, A.~J., \& {Scott}, P. 2009, \araa, 47,
  481, \dodoi{10.1146/annurev.astro.46.060407.145222}

\bibitem[{{Balona} {et~al.}(2015){Balona}, {Broomhall}, {Kosovichev},
  {Nakariakov}, {Pugh}, \& {Van Doorsselaere}}]{2015MNRAS.450..956B}
{Balona}, L.~A., {Broomhall}, A.~M., {Kosovichev}, A., {et~al.} 2015, \mnras,
  450, 956, \dodoi{10.1093/mnras/stv661}

\bibitem[{{Casagrande} {et~al.}(2007){Casagrande}, {Flynn}, {Portinari},
  {Girardi}, \& {Jimenez}}]{2007MNRAS.382.1516C}
{Casagrande}, L., {Flynn}, C., {Portinari}, L., {Girardi}, L., \& {Jimenez}, R.
  2007, \mnras, 382, 1516, \dodoi{10.1111/j.1365-2966.2007.12512.x}

\bibitem[{{Cho} {et~al.}(2016){Cho}, {Cho}, {Nakariakov}, {Kim}, \&
  {Kumar}}]{2016ApJ...830..110C}
{Cho}, I.~H., {Cho}, K.~S., {Nakariakov}, V.~M., {Kim}, S., \& {Kumar}, P.
  2016, \apj, 830, 110, \dodoi{10.3847/0004-637X/830/2/110}

\bibitem[{{Duckenfield} {et~al.}(2021){Duckenfield}, {Kolotkov}, \&
  {Nakariakov}}]{2021A&A...646A.155D}
{Duckenfield}, T.~J., {Kolotkov}, D.~Y., \& {Nakariakov}, V.~M. 2021, \aap,
  646, A155, \dodoi{10.1051/0004-6361/202039791}

\bibitem[{{Feldman} {et~al.}(1995){Feldman}, {Laming}, \&
  {Doschek}}]{1995ApJ...451L..79F}
{Feldman}, U., {Laming}, J.~M., \& {Doschek}, G.~A. 1995, \apjl, 451, L79,
  \dodoi{10.1086/309695}

\bibitem[{{Franciosini} {et~al.}(1999){Franciosini}, {Massi}, {Paredes}, \&
  {Estalella}}]{1999A&A...341..595F}
{Franciosini}, E., {Massi}, M., {Paredes}, J.~M., \& {Estalella}, R. 1999,
  \aap, 341, 595

\bibitem[{{G{\"u}del}(2004)}]{2004A&ARv..12...71G}
{G{\"u}del}, M. 2004, \aapr, 12, 71, \dodoi{10.1007/s00159-004-0023-2}

\bibitem[{{Hawley} \& {Pettersen}(1991)}]{1991ApJ...378..725H}
{Hawley}, S.~L., \& {Pettersen}, B.~R. 1991, \apj, 378, 725,
  \dodoi{10.1086/170474}

\bibitem[{{Kim} {et~al.}(2012){Kim}, {Nakariakov}, \&
  {Shibasaki}}]{2012ApJ...756L..36K}
{Kim}, S., {Nakariakov}, V.~M., \& {Shibasaki}, K. 2012, \apjl, 756, L36,
  \dodoi{10.1088/2041-8205/756/2/L36}

\bibitem[{{Klimchuk} {et~al.}(1992){Klimchuk}, {Lemen}, {Feldman}, {Tsuneta},
  \& {Uchida}}]{1992PASJ...44L.181K}
{Klimchuk}, J.~A., {Lemen}, J.~R., {Feldman}, U., {Tsuneta}, S., \& {Uchida},
  Y. 1992, \pasj, 44, L181

\bibitem[{{Kowalski} {et~al.}(2013){Kowalski}, {Hawley}, {Wisniewski}, {Osten},
  {Hilton}, {Holtzman}, {Schmidt}, \& {Davenport}}]{2013ApJS..207...15K}
{Kowalski}, A.~F., {Hawley}, S.~L., {Wisniewski}, J.~P., {et~al.} 2013, \apjs,
  207, 15, \dodoi{10.1088/0067-0049/207/1/15}

\bibitem[{{Kumar} {et~al.}(2013){Kumar}, {Innes}, \&
  {Inhester}}]{2013ApJ...779L...7K}
{Kumar}, P., {Innes}, D.~E., \& {Inhester}, B. 2013, \apjl, 779, L7,
  \dodoi{10.1088/2041-8205/779/1/L7}

\bibitem[{{Kupriyanova} {et~al.}(2020){Kupriyanova}, {Kolotkov}, {Nakariakov},
  \& {Kaufman}}]{2020STP.....6a...3K}
{Kupriyanova}, E., {Kolotkov}, D., {Nakariakov}, V., \& {Kaufman}, A. 2020,
  Solar-Terrestrial Physics, 6, 3, \dodoi{10.12737/stp-61202001}

\bibitem[{{Lenz} {et~al.}(1999){Lenz}, {DeLuca}, {Golub}, {Rosner}, \&
  {Bookbinder}}]{1999ApJ...517L.155L}
{Lenz}, D.~D., {DeLuca}, E.~E., {Golub}, L., {Rosner}, R., \& {Bookbinder},
  J.~A. 1999, \apjl, 517, L155, \dodoi{10.1086/312045}

\bibitem[{{L{\'o}pez-Santiago}(2018)}]{2018RSPTA.37670253L}
{L{\'o}pez-Santiago}, J. 2018, Philosophical Transactions of the Royal Society
  of London Series A, 376, 20170253, \dodoi{10.1098/rsta.2017.0253}

\bibitem[{{L{\'o}pez-Santiago} {et~al.}(2016){L{\'o}pez-Santiago},
  {Crespo-Chac{\'o}n}, {Flaccomio}, {Sciortino}, {Micela}, \&
  {Reale}}]{2016A&A...590A...7L}
{L{\'o}pez-Santiago}, J., {Crespo-Chac{\'o}n}, I., {Flaccomio}, E., {et~al.}
  2016, \aap, 590, A7, \dodoi{10.1051/0004-6361/201527499}

\bibitem[{{Mancuso} {et~al.}(2020){Mancuso}, {Barghini}, \&
  {Telloni}}]{2020A&A...636A..96M}
{Mancuso}, S., {Barghini}, D., \& {Telloni}, D. 2020, \aap, 636, A96,
  \dodoi{10.1051/0004-6361/201936819}

\bibitem[{{Mariska}(2005)}]{2005ApJ...620L..67M}
{Mariska}, J.~T. 2005, \apjl, 620, L67, \dodoi{10.1086/428611}

\bibitem[{{Mariska}(2006)}]{2006ApJ...639..484M}
---. 2006, \apj, 639, 484, \dodoi{10.1086/499296}

\bibitem[{{Mariska} {et~al.}(2008){Mariska}, {Warren}, {Williams}, \&
  {Watanabe}}]{2008ApJ...681L..41M}
{Mariska}, J.~T., {Warren}, H.~P., {Williams}, D.~R., \& {Watanabe}, T. 2008,
  \apjl, 681, L41, \dodoi{10.1086/590341}

\bibitem[{{McLaughlin} {et~al.}(2018){McLaughlin}, {Nakariakov}, {Dominique},
  {Jel{\'\i}nek}, \& {Takasao}}]{2018SSRv..214...45M}
{McLaughlin}, J.~A., {Nakariakov}, V.~M., {Dominique}, M., {Jel{\'\i}nek}, P.,
  \& {Takasao}, S. 2018, \ssr, 214, 45, \dodoi{10.1007/s11214-018-0478-5}

\bibitem[{{Mendoza-Brice{\~n}o} {et~al.}(2004){Mendoza-Brice{\~n}o},
  {Erd{\'e}lyi}, \& {Sigalotti}}]{2004ApJ...605..493M}
{Mendoza-Brice{\~n}o}, C.~A., {Erd{\'e}lyi}, R., \& {Sigalotti}, L. D.~G. 2004,
  \apj, 605, 493, \dodoi{10.1086/382182}

\bibitem[{{Mitra-Kraev} {et~al.}(2005){Mitra-Kraev}, {Harra}, {Williams}, \&
  {Kraev}}]{2005A&A...436.1041M}
{Mitra-Kraev}, U., {Harra}, L.~K., {Williams}, D.~R., \& {Kraev}, E. 2005,
  \aap, 436, 1041, \dodoi{10.1051/0004-6361:20052834}

\bibitem[{{Nakariakov} \& {Kolotkov}(2020)}]{2020ARA&A..58..441N}
{Nakariakov}, V.~M., \& {Kolotkov}, D.~Y. 2020, \araa, 58, 441,
  \dodoi{10.1146/annurev-astro-032320-042940}

\bibitem[{{Nakariakov} {et~al.}(2019{\natexlab{a}}){Nakariakov}, {Kolotkov},
  {Kupriyanova}, {Mehta}, {Pugh}, {Lee}, \& {Broomhall}}]{2019PPCF...61a4024N}
{Nakariakov}, V.~M., {Kolotkov}, D.~Y., {Kupriyanova}, E.~G., {et~al.}
  2019{\natexlab{a}}, Plasma Physics and Controlled Fusion, 61, 014024,
  \dodoi{10.1088/1361-6587/aad97c}

\bibitem[{{Nakariakov} {et~al.}(2019{\natexlab{b}}){Nakariakov}, {Kosak},
  {Kolotkov}, {Anfinogentov}, {Kumar}, \& {Moon}}]{2019ApJ...874L...1N}
{Nakariakov}, V.~M., {Kosak}, M.~K., {Kolotkov}, D.~Y., {et~al.}
  2019{\natexlab{b}}, \apjl, 874, L1, \dodoi{10.3847/2041-8213/ab0c9f}

\bibitem[{{Nakariakov} {et~al.}(2004){Nakariakov}, {Tsiklauri}, {Kelly},
  {Arber}, \& {Aschwanden}}]{2004A&A...414L..25N}
{Nakariakov}, V.~M., {Tsiklauri}, D., {Kelly}, A., {Arber}, T.~D., \&
  {Aschwanden}, M.~J. 2004, \aap, 414, L25, \dodoi{10.1051/0004-6361:20031738}

\bibitem[{{Nakariakov} \& {Zimovets}(2011)}]{2011ApJ...730L..27N}
{Nakariakov}, V.~M., \& {Zimovets}, I.~V. 2011, \apjl, 730, L27,
  \dodoi{10.1088/2041-8205/730/2/L27}

\bibitem[{{Ofman} \& {Wang}(2002)}]{2002ApJ...580L..85O}
{Ofman}, L., \& {Wang}, T. 2002, \apjl, 580, L85, \dodoi{10.1086/345548}

\bibitem[{{Pandey} \& {Srivastava}(2009)}]{2009ApJ...697L.153P}
{Pandey}, J.~C., \& {Srivastava}, A.~K. 2009, \apjl, 697, L153,
  \dodoi{10.1088/0004-637X/697/2/L153}

\bibitem[{{Pestalozzi} {et~al.}(2000){Pestalozzi}, {Benz}, {Conway}, \&
  {G{\"u}del}}]{2000A&A...353..569P}
{Pestalozzi}, M.~R., {Benz}, A.~O., {Conway}, J.~E., \& {G{\"u}del}, M. 2000,
  \aap, 353, 569.
\newblock \doarXiv{astro-ph/9912159}

\bibitem[{{Peter} \& {Bingert}(2012)}]{2012A&A...548A...1P}
{Peter}, H., \& {Bingert}, S. 2012, \aap, 548, A1,
  \dodoi{10.1051/0004-6361/201219473}

\bibitem[{{Peterson} {et~al.}(2010){Peterson}, {Mutel}, {G{\"u}del}, \&
  {Goss}}]{2010Natur.463..207P}
{Peterson}, W.~M., {Mutel}, R.~L., {G{\"u}del}, M., \& {Goss}, W.~M. 2010,
  \nat, 463, 207, \dodoi{10.1038/nature08643}

\bibitem[{{Pugh} {et~al.}(2016){Pugh}, {Armstrong}, {Nakariakov}, \&
  {Broomhall}}]{2016MNRAS.459.3659P}
{Pugh}, C.~E., {Armstrong}, D.~J., {Nakariakov}, V.~M., \& {Broomhall}, A.~M.
  2016, \mnras, 459, 3659, \dodoi{10.1093/mnras/stw850}

\bibitem[{{Pugh} {et~al.}(2015){Pugh}, {Nakariakov}, \&
  {Broomhall}}]{2015ApJ...813L...5P}
{Pugh}, C.~E., {Nakariakov}, V.~M., \& {Broomhall}, A.~M. 2015, \apjl, 813, L5,
  \dodoi{10.1088/2041-8205/813/1/L5}

\bibitem[{{Reale}(2014)}]{2014LRSP...11....4R}
{Reale}, F. 2014, Living Reviews in Solar Physics, 11, 4,
  \dodoi{10.12942/lrsp-2014-4}

\bibitem[{{Reale} {et~al.}(2018){Reale}, {Lopez-Santiago}, {Flaccomio},
  {Petralia}, \& {Sciortino}}]{2018ApJ...856...51R}
{Reale}, F., {Lopez-Santiago}, J., {Flaccomio}, E., {Petralia}, A., \&
  {Sciortino}, S. 2018, \apj, 856, 51, \dodoi{10.3847/1538-4357/aaaf1f}

\bibitem[{{Reale} {et~al.}(2019){Reale}, {Testa}, {Petralia}, \&
  {Kolotkov}}]{2019ApJ...884..131R}
{Reale}, F., {Testa}, P., {Petralia}, A., \& {Kolotkov}, D.~Y. 2019, \apj, 884,
  131, \dodoi{10.3847/1538-4357/ab4270}

\bibitem[{{Reiners}(2012)}]{2012LRSP....9....1R}
{Reiners}, A. 2012, Living Reviews in Solar Physics, 9, 1,
  \dodoi{10.12942/lrsp-2012-1}

\bibitem[{{Reznikova} \& {Shibasaki}(2011)}]{2011A&A...525A.112R}
{Reznikova}, V.~E., \& {Shibasaki}, K. 2011, \aap, 525, A112,
  \dodoi{10.1051/0004-6361/201015600}

\bibitem[{{Rojas-Ayala} {et~al.}(2012){Rojas-Ayala}, {Covey}, {Muirhead}, \&
  {Lloyd}}]{2012ApJ...748...93R}
{Rojas-Ayala}, B., {Covey}, K.~R., {Muirhead}, P.~S., \& {Lloyd}, J.~P. 2012,
  \apj, 748, 93, \dodoi{10.1088/0004-637X/748/2/93}

\bibitem[{{Selwa} {et~al.}(2005){Selwa}, {Murawski}, \&
  {Solanki}}]{2005A&A...436..701S}
{Selwa}, M., {Murawski}, K., \& {Solanki}, S.~K. 2005, \aap, 436, 701,
  \dodoi{10.1051/0004-6361:20042319}

\bibitem[{{Srivastava} \& {Dwivedi}(2010)}]{2010NewA...15....8S}
{Srivastava}, A.~K., \& {Dwivedi}, B.~N. 2010, \na, 15, 8,
  \dodoi{10.1016/j.newast.2009.05.006}

\bibitem[{{Srivastava} {et~al.}(2013){Srivastava}, {Lalitha}, \&
  {Pandey}}]{2013ApJ...778L..28S}
{Srivastava}, A.~K., {Lalitha}, S., \& {Pandey}, J.~C. 2013, \apjl, 778, L28,
  \dodoi{10.1088/2041-8205/778/2/L28}

\bibitem[{{Stepanov} {et~al.}(2006){Stepanov}, {Tsap}, \&
  {Kopylova}}]{2006AstL...32..569S}
{Stepanov}, A.~V., {Tsap}, Y.~T., \& {Kopylova}, Y.~G. 2006, Astronomy Letters,
  32, 569, \dodoi{10.1134/S1063773706080081}

\bibitem[{{Taroyan} {et~al.}(2005){Taroyan}, {Erd{\'e}lyi}, {Doyle}, \&
  {Bradshaw}}]{2005A&A...438..713T}
{Taroyan}, Y., {Erd{\'e}lyi}, R., {Doyle}, J.~G., \& {Bradshaw}, S.~J. 2005,
  \aap, 438, 713, \dodoi{10.1051/0004-6361:20052794}

\bibitem[{{Tsang} {et~al.}(2012){Tsang}, {Pun}, {Di Stefano}, {Li}, \&
  {Kong}}]{2012ApJ...754..107T}
{Tsang}, B.~T.~H., {Pun}, C.~S.~J., {Di Stefano}, R., {Li}, K.~L., \& {Kong},
  A.~K.~H. 2012, \apj, 754, 107, \dodoi{10.1088/0004-637X/754/2/107}

\bibitem[{{Tsiklauri} \& {Nakariakov}(2001)}]{2001A&A...379.1106T}
{Tsiklauri}, D., \& {Nakariakov}, V.~M. 2001, \aap, 379, 1106,
  \dodoi{10.1051/0004-6361:20011378}

\bibitem[{{Vida} {et~al.}(2019){Vida}, {Ol{\'a}h}, {K{\H{o}}v{\'a}ri}, {van
  Driel-Gesztelyi}, {Mo{\'o}r}, \& {P{\'a}l}}]{2019ApJ...884..160V}
{Vida}, K., {Ol{\'a}h}, K., {K{\H{o}}v{\'a}ri}, Z., {et~al.} 2019, \apj, 884,
  160, \dodoi{10.3847/1538-4357/ab41f5}

\bibitem[{{Wang} {et~al.}(2015){Wang}, {Ofman}, {Sun}, {Provornikova}, \&
  {Davila}}]{2015ApJ...811L..13W}
{Wang}, T., {Ofman}, L., {Sun}, X., {Provornikova}, E., \& {Davila}, J.~M.
  2015, \apjl, 811, L13, \dodoi{10.1088/2041-8205/811/1/L13}

\bibitem[{{Wang} {et~al.}(2021){Wang}, {Ofman}, {Yuan}, {Reale}, {Kolotkov}, \&
  {Srivastava}}]{2021SSRv..217...34W}
{Wang}, T., {Ofman}, L., {Yuan}, D., {et~al.} 2021, \ssr, 217, 34,
  \dodoi{10.1007/s11214-021-00811-0}

\bibitem[{{Wang} {et~al.}(2002){Wang}, {Solanki}, {Curdt}, {Innes}, \&
  {Dammasch}}]{2002ApJ...574L.101W}
{Wang}, T., {Solanki}, S.~K., {Curdt}, W., {Innes}, D.~E., \& {Dammasch}, I.~E.
  2002, \apjl, 574, L101, \dodoi{10.1086/342189}

\bibitem[{{Wang} {et~al.}(2003){Wang}, {Solanki}, {Curdt}, {Innes}, {Dammasch},
  \& {Kliem}}]{2003A&A...406.1105W}
{Wang}, T.~J., {Solanki}, S.~K., {Curdt}, W., {et~al.} 2003, \aap, 406, 1105,
  \dodoi{10.1051/0004-6361:20030858}

\bibitem[{{Watko} \& {Klimchuk}(2000)}]{2000SoPh..193...77W}
{Watko}, J.~A., \& {Klimchuk}, J.~A. 2000, \solphys, 193, 77,
  \dodoi{10.1023/A:1005209528612}

\bibitem[{{Welsh} {et~al.}(2006){Welsh}, {Wheatley}, {Browne}, {Siegmund},
  {Doyle}, {O'Shea}, {Antonova}, {Forster}, {Seibert}, {Morrissey}, \&
  {Taroyan}}]{2006A&A...458..921W}
{Welsh}, B.~Y., {Wheatley}, J., {Browne}, S.~E., {et~al.} 2006, \aap, 458, 921,
  \dodoi{10.1051/0004-6361:20065304}

\bibitem[{{Zaitsev} {et~al.}(2004){Zaitsev}, {Kislyakov}, {Stepanov}, {Kliem},
  \& {Furst}}]{2004AstL...30..319Z}
{Zaitsev}, V.~V., {Kislyakov}, A.~G., {Stepanov}, A.~V., {Kliem}, B., \&
  {Furst}, E. 2004, Astronomy Letters, 30, 319, \dodoi{10.1134/1.1738154}

\bibitem[{{Zavershinskii} {et~al.}(2019){Zavershinskii}, {Kolotkov},
  {Nakariakov}, {Molevich}, \& {Ryashchikov}}]{2019PhPl...26h2113Z}
{Zavershinskii}, D.~I., {Kolotkov}, D.~Y., {Nakariakov}, V.~M., {Molevich},
  N.~E., \& {Ryashchikov}, D.~S. 2019, Physics of Plasmas, 26, 082113,
  \dodoi{10.1063/1.5115224}

\bibitem[{{Zimovets} {et~al.}(2021){Zimovets}, {McLaughlin}, {Srivastava},
  {Kolotkov}, {Kuznetsov}, {Kupriyanova}, {Cho}, {Inglis}, {Reale}, {Pascoe},
  {Tian}, {Yuan}, {Li}, \& {Zhang}}]{2021SSRv..217...66Z}
{Zimovets}, I.~V., {McLaughlin}, J.~A., {Srivastava}, A.~K., {et~al.} 2021,
  \ssr, 217, 66, \dodoi{10.1007/s11214-021-00840-9}

\end{thebibliography}
\bibliographystyle{aasjournal}

\end{document}